\documentstyle[footmisc,gcdv,psfig]{article}

\def\reference{\parskip 0pt\par\noindent\hangindent 0.5 truecm}
\def\spose#1{\hbox to 0pt{#1\hss}}
\def\simlt{\mathrel{\spose{\lower 3pt\hbox{$\mathchar"218$}}
     \raise 2.0pt\hbox{$\mathchar"13C$}}}
\def\simgt{\mathrel{\spose{\lower 3pt\hbox{$\mathchar"218$}}
     \raise 2.0pt\hbox{$\mathchar"13E$}}}
\def\eg{{\rm e.g.}}

\def\etal{{\rm et~al.}}   
\def\GCD+{{\sf GCD+}}
\def\HI{H{\small\sc{I}}}
\def\e#1{\ifmmode{\times 10^{#1}}\else$ \times 10^{#1}$\fi}

\def\msun{\ifmmode{{\rm \,M}_\odot}\else${\rm \,M}_\odot$\fi}
\def\kms{\ifmmode{{\rm \,km\,s}^{-1}}\else${\rm \,km\,s}^{-1}$\fi}
\def\Gyr{\ifmmode{{\rm \,Gyr}}\else${\rm \,Gyr}$\fi}
\def\Myr{\ifmmode{{\rm \,Myr}}\else${\rm \,Myr}$\fi}
\def\kpc{\ifmmode{{\rm \,kpc}}\else${\rm \,kpc}$\fi}
\def\Jybeam{\ifmmode{{\rm \,Jy/beam}}\else${\rm \,Jy/beam}$\fi}
\def\vlsr{\ifmmode{{\rm v}_{\rm LSR}}\else${\rm v}_{\rm LSR}$\fi}
\def\vgsr{\ifmmode{{\rm v}_{\rm GSR}}\else${\rm v}_{\rm GSR}$\fi}
\def\deg{\ifmmode^{\circ}\else$^{\circ}$\fi}

\begin{document}

\small
\shorttitle{N-body Simulations of the Magellanic Stream}
\shortauthor{T.W. Connors, D. Kawata, S.T. Maddison \& B.K. Gibson}

\title{\large \bf
  High-resolution N-body Simulations of Galactic Cannibalism: \\ 
  The Magellanic Stream
}

\author{\small
  Tim W Connors$^{1}$,
  Daisuke Kawata$^{1}$,
  Sarah T. Maddison$^{1}$,
  Brad K. Gibson$^{1}$
}

\date{}
\twocolumn[
\maketitle
\vspace{-20pt}
\small
{\center
  Centre for Astrophysics \& Supercomputing, 
  Swinburne University, Hawthorn, VIC 3122, 
  Australia\\
  $^1$tconnors,dkawata,bgibson,smaddiso@astro.swin.edu.au\\[3mm]
}

\begin{center}
{\bfseries Abstract}
\end{center}
\begin{quotation}
\begin{small}
\vspace{-5pt}
  Hierarchical clustering represents the favoured paradigm for galaxy
  formation throughout the Universe; due to its proximity, the
  Magellanic system offers one of the few opportunities for
  astrophysicists to decompose the full six-dimensional phase-space
  history of a satellite in the midst of being cannibalised by its
  host galaxy. The availability of improved observational data for the
  Magellanic Stream and parallel advances in computational power has
  led us to revisit the canonical tidal model describing the
  disruption of the Small Magellanic Cloud and the consequent
  formation of the Stream.  We suggest improvements to the tidal model
  in light of these recent advances.
\end{small}
\end{quotation}
]

\bigskip
\bigskip

\section{Introduction}

With the release of the Wilkinson Microwave Anisotropy Probe (WMAP)
results (Spergel et~al. 2003), the Cold Dark Matter (CDM) has
essentially shifted from a ``favoured'' paradigm to what is now
referred to as the ``concordance model''.  Hierarchical clustering is
an important component of CDM models, one in which the first objects
to collapse in the Universe were small, with subsequent merging of
these objects coupled with collapse on increasingly larger scales as
the Universe ages.  Such merger- and accretion-driven evolution
appears to have peaked over the redshift range $\sim 2$ -- 5 (e.g.
Murali et~al. 2002), but equally important, continues to the
present-day.  Indeed, our own Local Group provides several spectacular
examples of hierarchical clustering ``in action'', including the
disrupting Sagittarius dwarf (Ibata et~al. 1994), the putative Canis
Major dwarf (Martin et~al. 2003), and perhaps the most visually
stunning of all, the debris associated with the interacting Large and
Small Magellanic Clouds (LMC and SMC, respectively, hereafter) -- the
so-called Magellanic Stream (Mathewson et~al 1974).  Disrupting
satellites such as these are the best local laboratory to understand
the physical processes of ``galactic cannibalism'', as we have the
luxury of obtaining detailed observations pertaining to the respective
systems' star formation histories (e.g. Harris~\& Zaritsky 2001;
Smecker-Hane \etal\ 2002) and internal chemical evolution via stellar
abundance patterns for individual stars within the satellites (Tolstoy
et~al. 2003).

One of the most obvious of manifestations of cannibalism within the
Local Group is that of the aforementioned Magellanic Stream. The
Magellanic Stream (MS) is a remarkably colinear band of (primarily)
neutral hydrogen (\HI{}) stretching from horizon-to-horizon through
the South Galactic Pole, emanating from the Magellanic System.
van~Kuilenburg (1972)\footnote{Anomalously high-velocity gas features
near the South Galactic Pole had actually been known since the work of
Dieter (1965), but the link to the Magellanic System was not fully
appreciated until that of Mathewson et~al. (1974).} discovered a
lengthy high-velocity gas stream near the South Galactic Pole, while
Wannier \& Wrixon (1972) noted that the feature had a large and
smoothly varying velocity (from $\vlsr \sim 0$ to $-400 \kms$, or
$\vgsr \sim 0$ to $-200 \kms$), and was over $60\deg$ long (but only
$\sim 4\deg$ wide).  Mathewson et~al.  (1974) finally confirmed the
connection between this feature and the Magellanic Clouds (MCs),
suggesting the stream was 180\deg{} in length, lying on a great
circle. They showed the stream was clumpy, and gave the designations
MSI--VI to the six dominant clumps (Mathewson et~al.  1977).  Most
recently, Putman \etal\ (1998) showed what is now considered to be the
full extent of the stream, with the identification of a leading arm
feature (LAF) definitively associated with the Magellanic System.
Indirect supporting evidence for both the trailing and leading arm
streams being associated directly with the disrupting Magellanic
Clouds is also provided by the similarity in chemical ``fingerprints''
between the gas in the streams and the gas in the Clouds (Lu
et~al. 1998; Gibson et~al. 2000).

Building upon the seminal work of Murai \& Fujimoto (1980), recent
observational and theoretical analyses are consistent with the
suggestion that the Clouds are close to peri-Galacticon. For example,
the Galactocentric radial velocities of the Clouds are small at $84
\kms$ (van der Marel \etal\ 2002) and $7 \kms$ (Hardy, Suntzeff \&
Azzopardi 1989; Gardiner, Sawa \& Fujimoto 1994 -- hereafter GSF94)
for the LMC and SMC, respectively, compared with their respective
transverse velocities in the Galactocentric frame of $280 \kms$ (van
der Marel \etal\ 2002) and $\sim 200 \kms$ (Lin, Jones \& Klemola
1995) consistent with this hypothesis.  The closest approach to date,
both between the Clouds and between the MCs and the Milky Way occurred
$\sim 200 \Myr$ ago, and the orbital period of the SMC about the LMC
is $\sim 900 \Myr$, with the Clouds as a pair orbiting the Galaxy with
a period of order $1.5 \Gyr$ (\eg{} GSF94).  Early models from Murai
\& Fujimoto (1980) and Lin \& Lynden-Bell (1982), supplemented with
the recent proper motion work from Jones, Klemola \& Lin (1994) have
provided us with an accurate representation of the present-day orbital
characteristics of the Magellanic System.  The determination of the
orbital sense of the system demonstrates clearly that the MS is an
extension stretching beyond the present galactocentric distance of the
MCs, rather than a bridge joining the MCs with the Galaxy, and leading
them (Lin \& Lynden-Bell 1982; Lin \etal\ 1995), and also that the MCs
are close to peri-Galacticon.

Considerable debate exists within the literature as to whether the
Magellanic Stream is the result of ram pressure stripping (Moore \&
Davis 1994; Mastropietro \etal\ 2004) or gravitational tidal effects
in which the Stream material is either stripped off the LMC (Weinberg
2000), the SMC (\eg{} Gardiner \& Noguchi 1996 -- hereafter GN96), or a
common envelope (the inter-Cloud region; \eg{} Heller \& Rohlfs 1994).
The observations of the LAF (Putman \etal\ 1998) show that tidal
forces account for at least some fraction of the ``force'' shaping the
existence of the Stream, even as the observed H$\alpha$ emission
measured along the Stream suggest that some additional ram pressure
heating effects may be present (Weiner \& Williams 1996; cf. Putman
et~al. 2003b).

Yoshizawa \& Noguchi (2003; hereafter YN03) have provided recently a
significant improvement to the now canonical ``tidal'' model of GN96,
via the inclusion of gas dynamics and star formation. In a prescient
forebearer to YN03, Gardiner (1999) also provided important extensions
to his earlier GN96 work using new constraints introduced by the
recent discovery of the LAF, the addition of a drag term into the
particle force equations, and an improved modelling of the LMC's disk
potential. These latter modifications have the beneficial effect of
mildly deflecting the orientation of the LAF with respect to the
Magellanic System in a manner more consistent with the Putman
et~al. (1998) dataset.

Encouraged by the success of these earlier studies, we are undertaking
a comprehensive computational program aimed at providing the
definitive deconstruction of this Rosetta Stone of hierarchical
clustering -- the disrupting Magellanic System.  We now have access to
the full HIPASS South {\it and } North dataset, data which was not
available to Putman et~al. (1998), allowing us to improve upon the
observational constraints on both the trailing Stream and leading arm.
Our {\it ultimate} product will be the construction of a model which
includes all relevant physical processes, including gas dynamics, ram
pressure, radiative cooling, star formation, and chemical enrichment,
all treated {\it self-consistently} for the first time, in a hope to
understand the physical processes of galactic cannibalism.  Our
cosmological chemodynamical code {\tt GCD+} (Kawata \& Gibson 2003a,b)
affords the power and flexibility to attack this problem in a manner
previously inaccessible.

What follows represents the first of a series of papers devoted to
this system; this Paper~I shows preliminary results based solely upon
very high-resolution N-body simulations undertaken without the gas
component of {\tt GCD+} implemented.  This first step was required in
order to allow a full exploration of orbital parameter space prior to
the introduction of gas into the modelling.  The reason for doing so
is that current observational constraints on the system still allow
one some flexibility in choosing a unique orbital configuration for
the system, partly due to our less-than-optimal understanding of the
LMC and SMC masses.  The spatial orientation and nature of any SMC
disk is also poorly constrained. Since N-body simulations are less
computationally ``expensive'', we can survey different orbits for the
Clouds, and determine our best orbital configuration(s).  In what
follows we present our current best N-body model for the Magellanic
Stream, compare this model with the extant observational data, and
provide a roadmap for our future work, highlighting the successes and
failures of the currently accepted canonical tidal model for the
formation of the Magellanic Stream.

\section{Simulations}

The basic framework of both GN96 and YN03 was adopted in our
study. The Milky Way (MW) and LMC were represented as fixed potentials
-- the MW with a flat rotation curve of $220 \kms$, and the LMC as a
Plummer potential with core radius of $3 \kpc$ and a mass of $2 \e{10}
\msun$.  Canonical wisdom suggests that the Stream results mainly from
the tidal disruption of the SMC (e.g. GSF94; GN96; Maddison
et~al. 2002; YN03), with minimal contribution from the LMC, and is
traditionally invoked (as we have done here) as a reasonable
justification for this assumption (cf. Mastropietro et~al. 2004).  The
SMC was modelled as a self-gravitating system of particles.  The
orbits for both the LMC and SMC were pre-calculated (see GN96 for
details), as derived by GSF94. The models were computed from $T_i \leq
-2 \Gyr$ to the present epoch ($T=0$), but note in passing that the
results are not dependent upon the specific starting epoch (GSF94).

We performed an extensive parameter search over those variables which
remain poorly-constrained by observation.  For the LMC and SMC, we
varied the ratio of the halo and disk masses between 1:1 and 5:3, the
ratio of the tidal radius of the halo-to-disk truncation radius
(albeit the ratio was restricted to values near unity), the scale
height of the disk (retaining a spherical halo with only marginal
deviations from sphericity), the velocity dispersion in the disk, and
the total masses of both clouds.  For the LMC, the mass range sampled
was 8 -- $20 \e9 \msun$ (Schommer
\etal\ 1992; Kunkel \etal\ 1997; van~der~Marel \etal\ 2002). The
parameters that most affected the position and quality of the
simulated Stream were the initial angle of the SMC disk relative to
the MW (surveyed over 2 dimensions on a 45\deg{} grid), and the radius
of the (non-stellar) disk (varied over radii between 2 and $7 \kpc$,
on a $0.5 \kpc$ grid).  We also performed convergence tests, both on
the number of particles and the starting epoch $T_i$.  Full results of
{\it all} tests will be the subject of Paper~II; for brevity, we only
present our preferred N-body model (based on kinematic and spatial
similarities to the observational data for the MS and LAF) here.

\begin{figure*}[t]
  \begin{center}
    \begin{tabular}{cc}
      \psfig{file=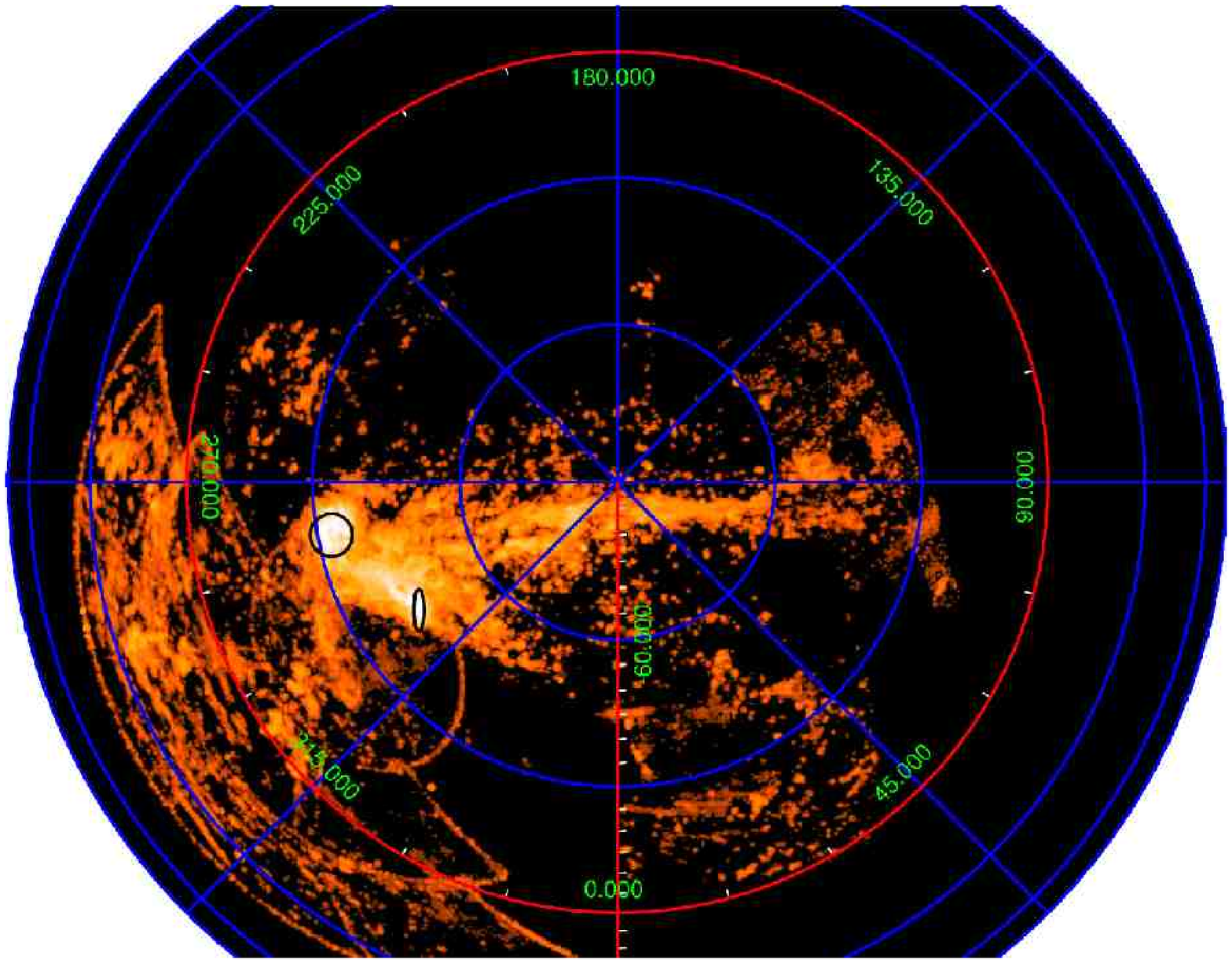,width=8cm} &
      \psfig{file=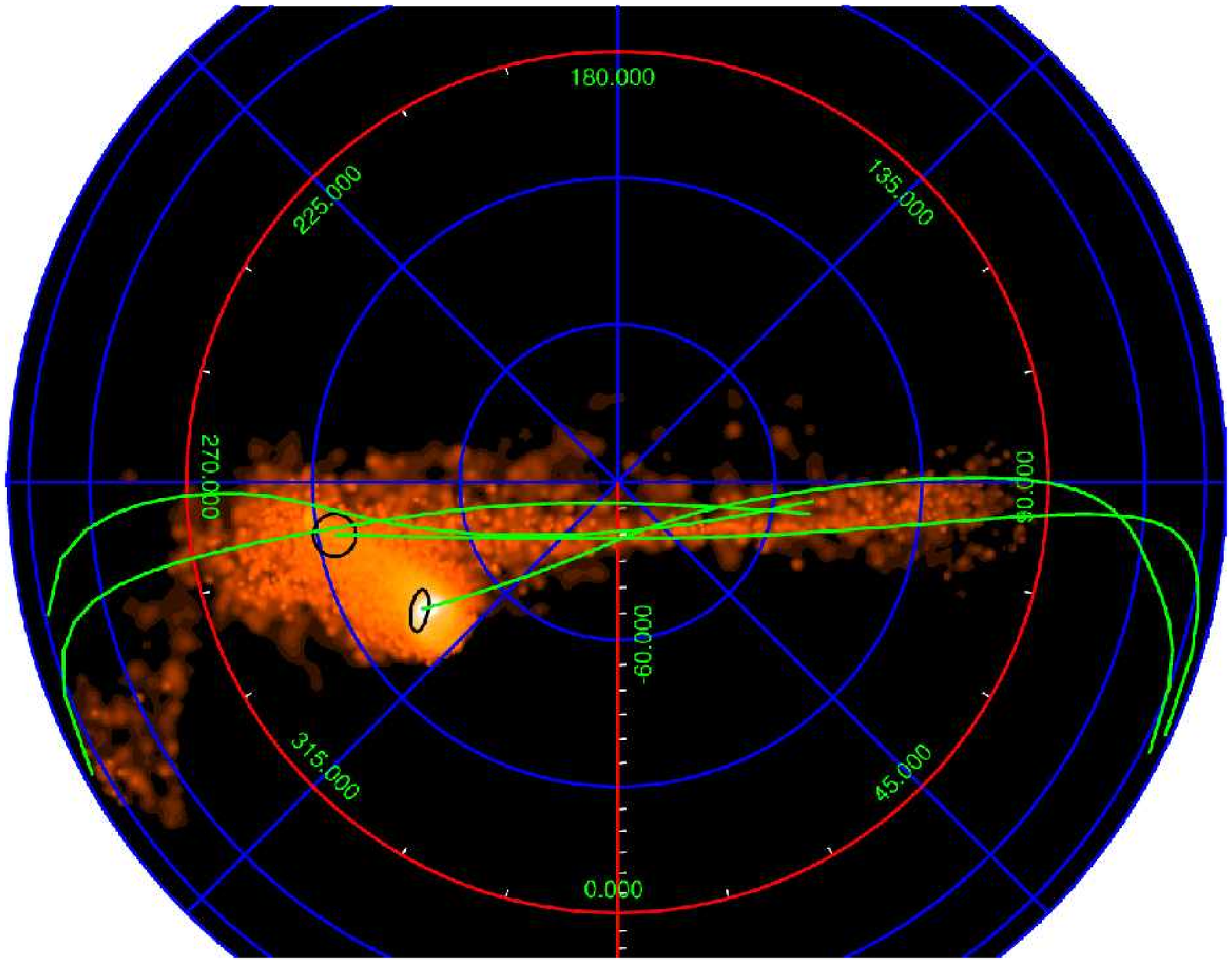,width=8cm}
    \end{tabular}

    \caption{Observed (left) and modelled (right) flux of the
    Magellanic Stream (trailing to the right of the Clouds in these
    panels) and LAF (to the left) on a logarithmic scale, in units of
    $\log_{10}(\rm {Jy/beam} \cdot \kms)$ -- peak flux of $1000
    \Jybeam \cdot \kms$, and minimum flux of $0.01 \Jybeam \cdot
    \kms$, on an all sky Zenith Equal Area projection.  Since only the
    SMC has been modelled, the flux around the LMC is underestimated.
    The projection of the initial SMC disk at the angle used in the
    simulation has been overlaid on the simulation at \mbox{$(l,b)
    \sim (300\deg,-45\deg)$}, as has the approximate size of the
    Plummer core radius of the LMC at \mbox{$(l,b) \sim
    (280\deg,-30\deg)$}.  The orbit calculated from the best model
    parameters has also been plotted in the right panel. The LAF
    (Putman et~al. 1998) is seen to extend from \mbox{$(l,b) \sim
    (290\deg,-30\deg)$} down to \mbox{$(l,b) \sim (310\deg,0\deg)$}
    and then back up to \mbox{$(l,b) \sim (290\deg,20\deg)$} and
    \emph{possibly} onto \mbox{$(l,b) \sim (265\deg,20\deg)$} in the
    left panel, and from \mbox{$(l,b) \sim (280\deg,-10\deg)$} to
    \mbox{$(l,b) \sim (305\deg,40\deg)$} and then to \mbox{$(l,b) \sim
    (290\deg,50\deg)$} in the right. The inter-Cloud bridge joins the
    two clouds and forms a common envelope around them. The Magellanic
    Stream is seen twisting to the right of the clouds, with many
    small clumps separated from the main stream in both panels.}
    \label{fig:skymom0} \end{center}
\end{figure*}

\begin{figure*}[t]
  \begin{center}
    \begin{tabular}{cc}
      \psfig{file=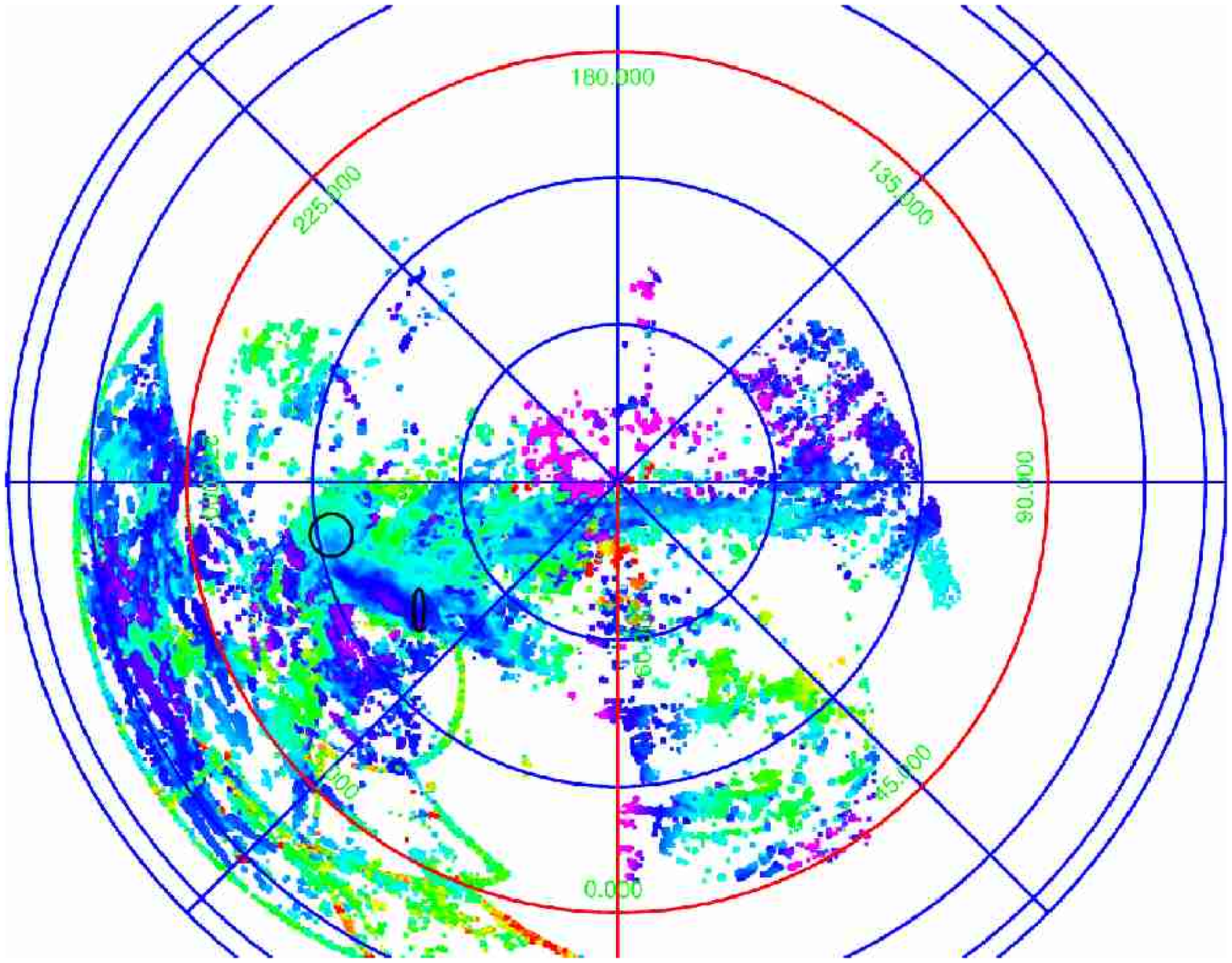,width=8cm} &
      \psfig{file=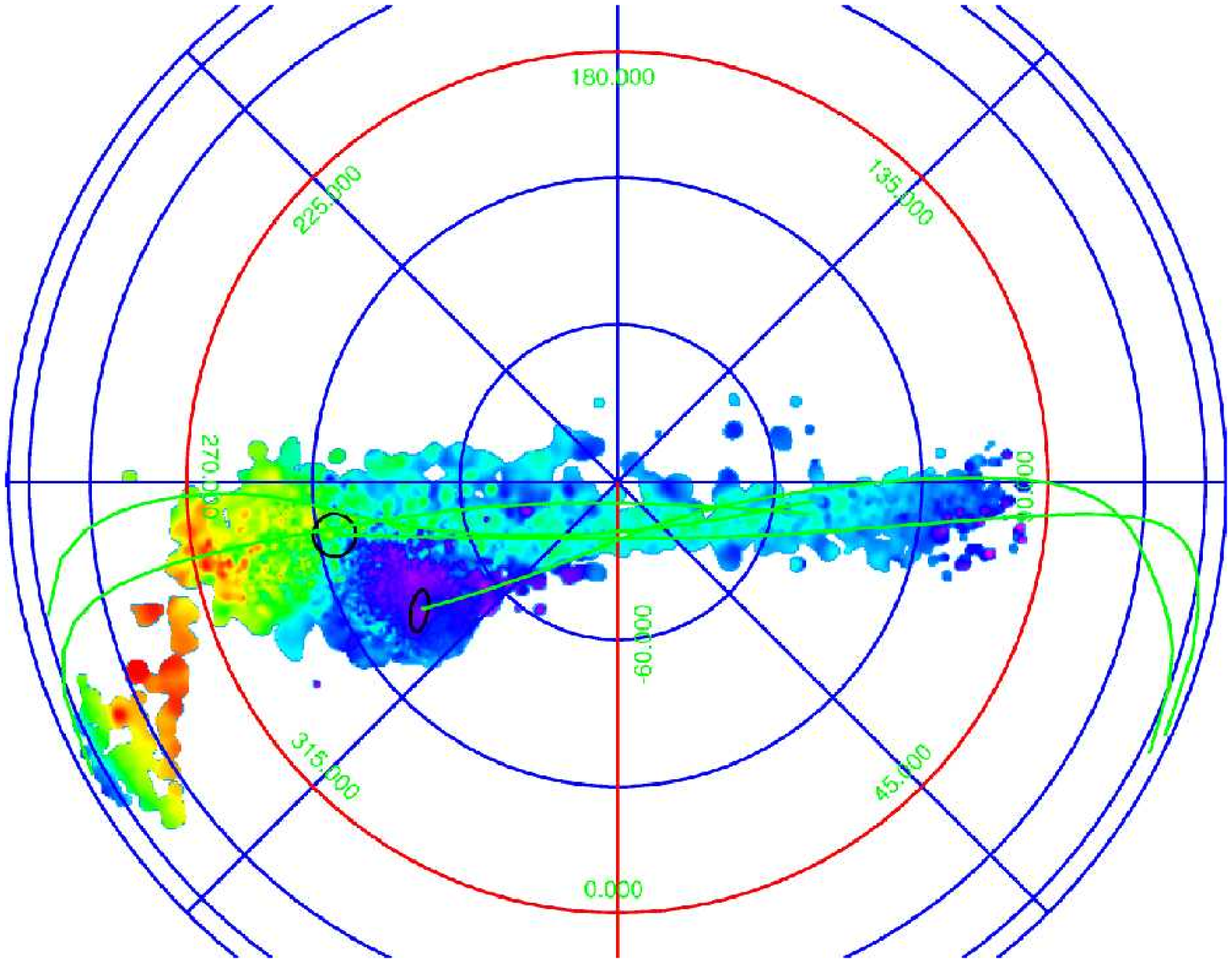,width=8cm}
    \end{tabular}
    \caption{As in Figure~\ref{fig:skymom0}, but now showing the first
      moment (with the mean velocity gradient subtracted equally from
      both images). The limiting column density employed in the
      simulation data is $2\e{17}{\rm cm}^{-2}$. The velocity 
      scale ranges from $-100 \kms$ (purple) to $+200 \kms$ (red).}
    \label{fig:skymom1}
  \end{center}
\end{figure*}

The initial particle distribution of our SMC disk was generated using
a modified version of {\tt galactICs} (Kuijken \& Dubinski 1995); a
two-component disk$+$halo model was the result of this first stage.
These particles were then evolved using the \GCD+ parallel tree N-body
code described by Kawata \& Gibson (2003a,b).  As noted earlier, the
preliminary work presented here was undertaken using only the N-body
component of \GCD+; full gas dynamics, star formation, cooling, and
chemical evolution will be explored in Paper~III.  In our initial
low-resolution runs we typically used 20000 disk particles and 33000
halo particles (to maintain an equal mass between disk and halo
particles).  We emphasise though that our high-resolution runs were
undertaken at a resolution ten times higher, corresponding to a
resolution $\sim 40$ times greater than that employed by GN96 and
YN03.  Such resolution allowed us to examine features of the MS, LAF,
and SMC, in a manner not previously possible, since smaller fractional
differences in particle density become statistically significant.  We
performed stability tests on the initial SMC particle configurations
removed from the influence of the potentials of both the host MW and
LMC.  The disk and halo were evolved together for $2 \Gyr$ and this
``relaxed'' particle distribution input into the MS simulation proper.

We found an improved match to the observational data when starting our
simulations at $T_i = -3 \Gyr$, marginally earlier than the $T_i = -2
\Gyr$ typically adopted (GN96; Gardiner 1999; YN03).  Our initial SMC
disk was slightly larger than that used by GN96 ($5.5 \kpc$ vs $5.0
\kpc$); in addition, our SMC disk mass was marginally less massive
($1.125 \e9 \msun$ vs $1.5 \e9 \msun$; with the total mass of the SMC
being $3 \e9 \msun$).  Finally, since the initial angles were
calculated on a regular grid, the chosen angle was slightly different
here -- \mbox{$(\theta, \phi) = (45\deg, 225\deg)$}, instead of
\mbox{$(\theta, \phi) = (45\deg, 230\deg)$}. The LMC was surveyed over
8, 10, 15 and $20 \e{9} \msun$, with the latter being favoured for the
choice of grid parameters adopted.

\section{Results}

Consistent with earlier models (e.g. GN96; YN03), an encounter between
the MW and the Clouds $1.5 \Gyr$ ago drew out the tidal features that
later became the LAF and MS under the tidal forces of the Galaxy (most
of the LAF material was pulled back into the inter-Cloud region by the
LMC). A stronger interaction between the LMC and SMC $\sim 200 \Myr$
ago resulted in an inter-Cloud Magellanic Bridge that has not yet had
time to disperse.

Assuming an \HI{} gas fraction of 0.76, and a conversion factor
between simulated column density in units of $\rm{atoms} \cdot
\rm{cm}^{-2}$ and \HI{} flux in units of \mbox{$\rm{Jy/beam} \cdot
\kms$}, of \mbox{$0.76 \times 1/(0.8 \times 1.823\e{18})$} (Barnes
et~al. 2001), Figure~\ref{fig:skymom0} shows both the observed \HI{}
flux of the Magellanic Stream and the simulated \HI{} flux for our
best model.  We see that the gross features of the Stream are
reproduced and conclude (as previous workers have) that the LAF
appears as a consequence of tidal interactions.  A failing of the
model lies in the discrepancy between the exact projected positions of
the observed and simulated LAFs, primarily in relation to the
respective points from which they appear to ``emanate''.  The actual
angle of deflection in both panels is quite similar, but the observed
``bend'' in the LAF back towards the great circle from \mbox{$(l,b)
\sim (310\deg,0\deg)$} to \mbox{$(l,b) \sim (290\deg,20\deg)$} and
\emph{possibly} onto \mbox{$(l,b) \sim (265\deg,20\deg)$} is not
reproduced. The increase in resolution of our best model over the
previous models allow us to see more detail in the LAF. We see there
is possibly a small ``kink'' back towards the great circle from
\mbox{$(l,b) \sim (305\deg,40\deg)$} to \mbox{$(l,b) \sim
(290\deg,50\deg)$}, that was not apparent in low resolution runs with
the same initial conditions. The definition of the LAF is somewhat
ambiguous in the model (particularly in the delineation between LAF
and SMC gas), but nevertheless, we find that the mass ratio between
our simulated MS and LAF is $\sim 5$.  The observed {\it mass} ratio
is difficult to quantify as one has no {\it direct} measure of the gas
mass in either the MS or LAF -- one is restricted to the \HI{} {\it
flux} ratio (Putman et~al. 1998).  Under the assumption that the MS
and LAF are at a comparable distance, the observed \HI{} gas mass
ratio is $\simgt 10$ (Putman et~al. 1998; Putman 2000).  The simulated
MS shown in Figure~\ref{fig:skymom0} is comparable in mass with that
observed -- assuming (i) a Stream distance of $50 \kpc$ and (ii) an
uncertain conversion from N-body to gas particle mass -- the simulated
MSI--MSVI clumps correspond to a mass of \mbox{$\sim 2.5 \e7 \msun$},
within a factor of four of the inferred empirical gas mass (Putman
et~al. 2003).

\begin{figure*}[t]
  \begin{center}
    \begin{tabular}{ll}
      \psfig{file=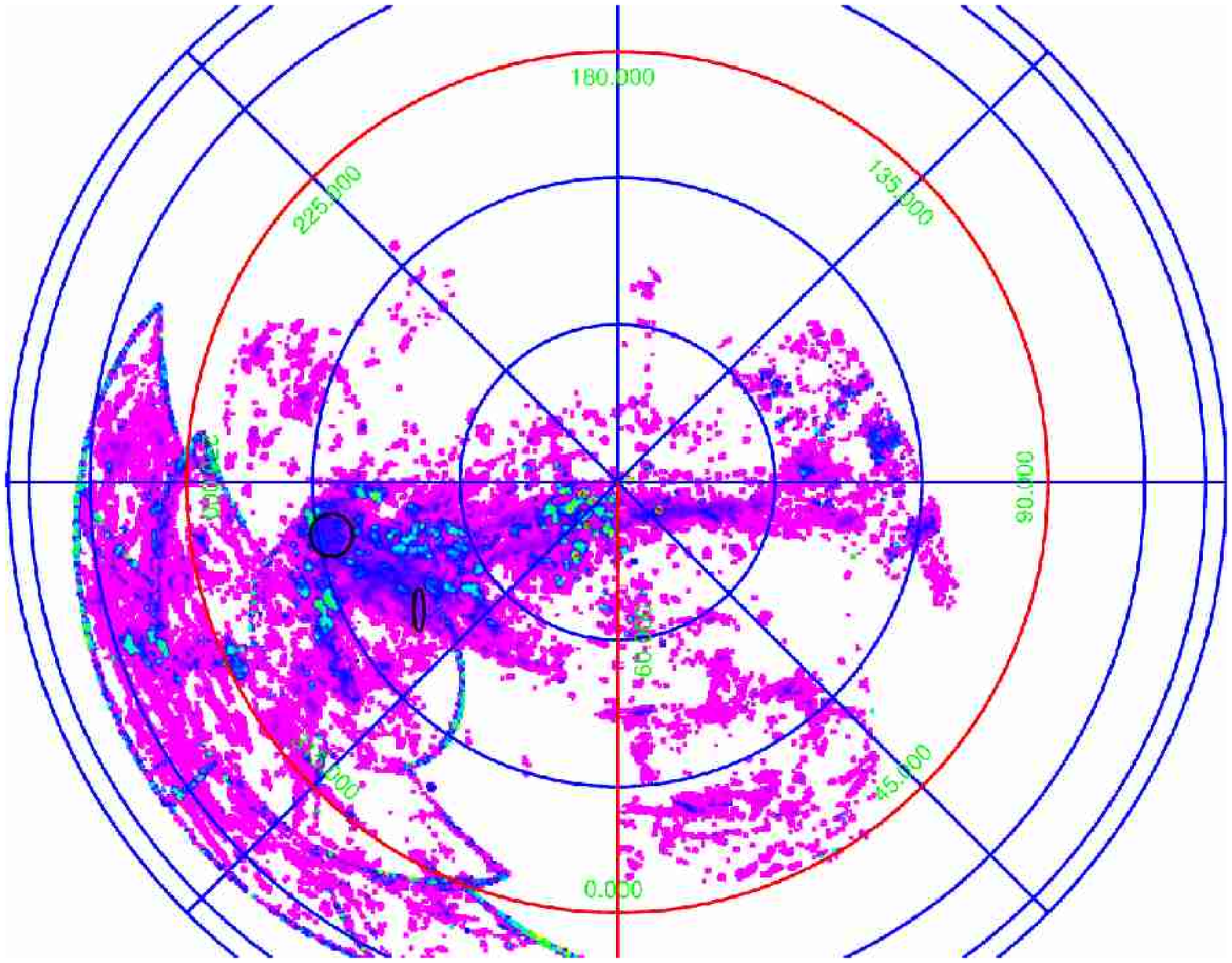,width=8cm} &
      \psfig{file=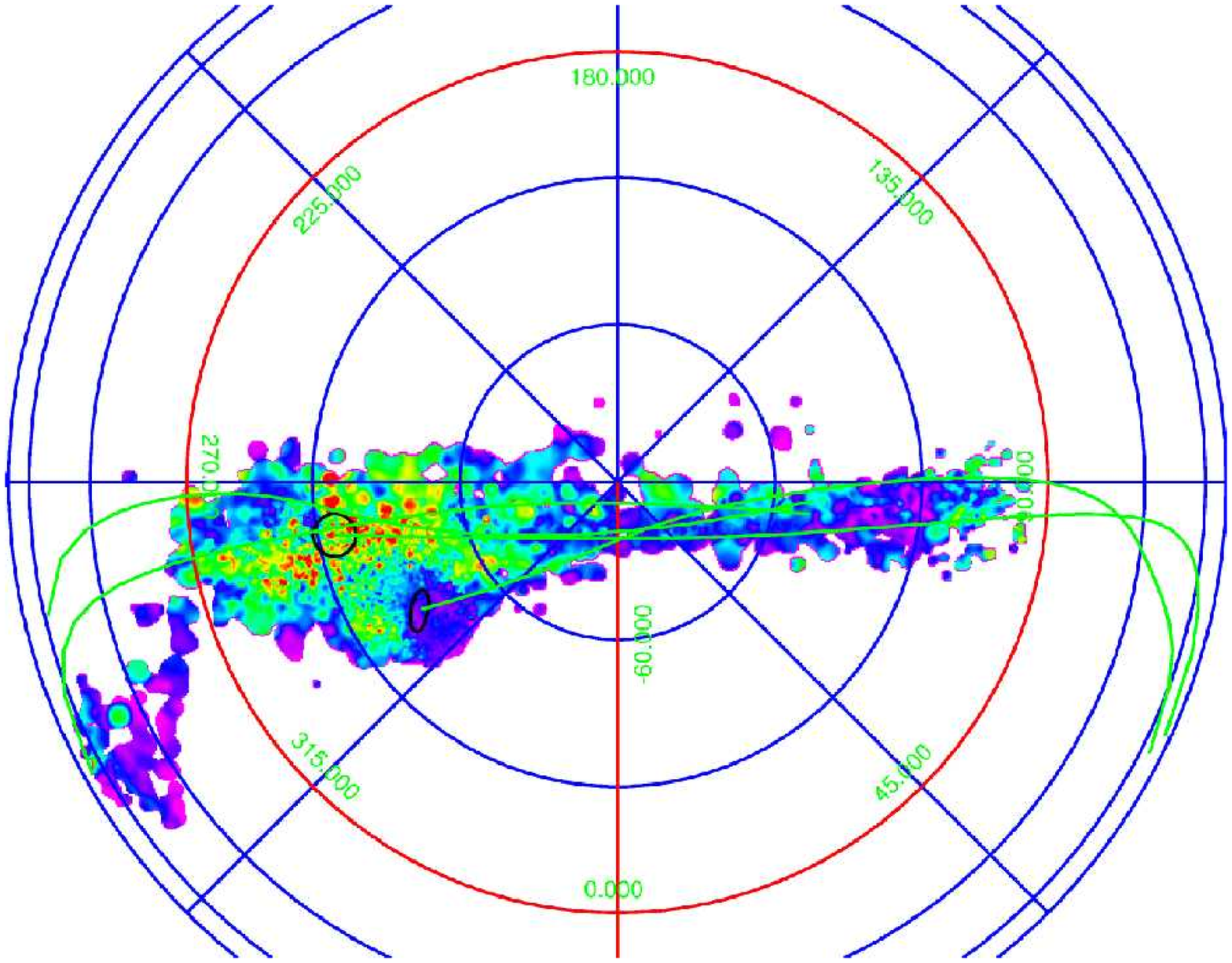,width=8cm}
    \end{tabular}
    \caption{As in Figure~\ref{fig:skymom1}, but now showing the second
      moment. The limiting column density employed in the simulation
      data is as in Figure~\ref{fig:skymom1}. The velocity dispersion
      ranges from from $10 \kms$ (purple) to $90 \kms$ (red).}
    \label{fig:skymom2}
  \end{center}
\end{figure*}

\begin{figure}[t]
  \begin{center}
    \psfig{file=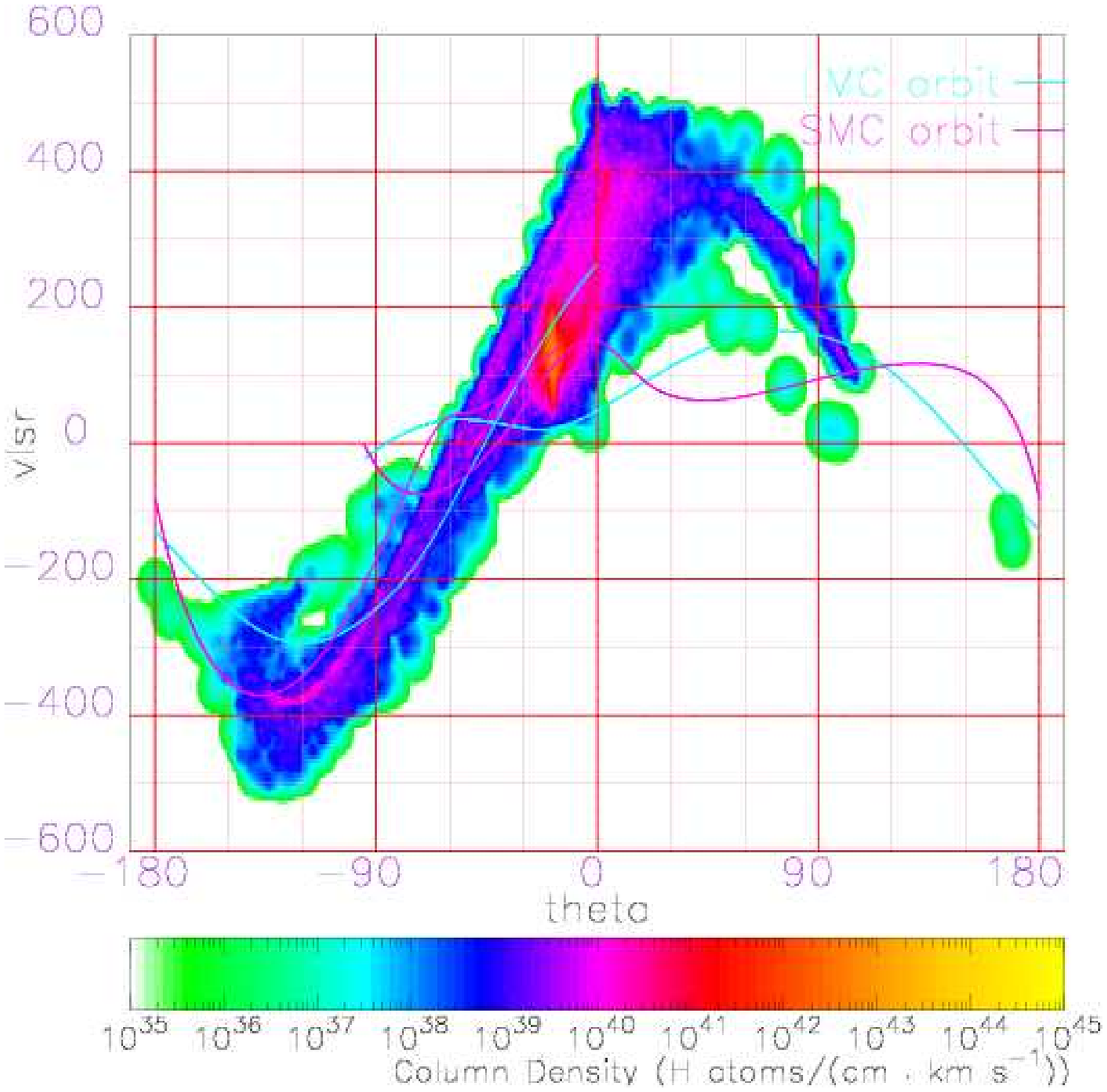,width=8cm}
    \caption{A ``column density'' plot of velocity relative to the
    local standard of rest (in \kms) for our preferred N-body model
    plotted against Magellanic Longitude $\theta$, where the
    ``density'' is in units of $\log_{10}(\rm{\HI{}\ atoms} / (\rm {cm}
    \cdot \kms))$.  The current LMC position defines the origin,
    $\theta = 0$ in this system. The orbit of the two Clouds are
    overlaid.  The MS (LAF) extends to negative (positive) $\theta$ in
    this projection.}  \label{fig:vlsrtheta} \end{center}
\end{figure}

Figure~\ref{fig:skymom1} shows the first moment map of both the
observed (left panel) and simulated (right) Streams after subtraction
of the observed gross velocity gradient -- i.e., the velocity
``residuals'' with respect to the smooth underlying
gradient.\footnote{By fitting two Fourier components to the observed
MS and LAF, as a function of Magellanic longitude (as defined in
Wannier \& Wrixon 1972).}  Any differences in the geometries between
the simulated and observed Streams can have a significant impact on
the velocity maps, due to the (a) reference frame transformation from
both the simulation ``cube'' and observations to the Galactic Standard
of Rest, and (b) the position-dependency of the velocity gradient
subtraction (see Figure~\ref{fig:vlsrtheta} for the overall trend with
position in the simulation). With this in mind, we see that there is
surprisingly good agreement between data and model over the entire MS
and LAF in velocity space.  Figure~\ref{fig:skymom2} shows the second
moment maps. The model admittedly does not reproduce the velocity
dispersion in the inter-Cloud region very well (the MS and LAF are in
better agreement though).  This discrepancy {\it may} be solved by the
inclusion of an appropriate treatment of gas dynamics (due to the
dissipative nature of the gas); this will be explored in Paper~III of
this series.

It is worth drawing attention to the observed bifurcation and twisting
of the Stream described in some detail by Morras (1983).  Putman
et~al. (2003) claim that this spatial bifurcation results from the
binary motion of the SMC around the LMC (Putman \etal\ 2003).  In this
picture, the filaments of the bifurcation are associated with gas
stripped from the SMC and inter-Cloud region of the LMC--SMC system.
In our simulations, this ``helix-like'' twisting of the filaments is a
natural consequence of the SMC--LMC orbits ``twisting'' about each
other -- the orbital overlays of Figure~\ref{fig:skymom0} (right
panel) show the near one-to-one correlation between observed filaments
locations and the projected orbits of the Clouds. This spatial
bifurcation (and a kinetic bifurcation seen in
Figure~\ref{fig:vlsrtheta}) are obvious only in our high resolution
simulation.

The preliminary simulations described here also yield a velocity
bifurcation of $\sim 100 \kms$ along the first $\sim 40 \deg$ of the
MS trailing the SMC (\mbox{$-50\deg < \theta < -10\deg$}).
Figure~\ref{fig:vlsrtheta} shows the velocity in the Local Standard of
Rest plotted against the Magellanic longitude (where $\theta$ is
defined with a somewhat different origin to that adopted by Wannier \&
Wrixon 1972 and Mathewson \etal\ 1974). A kinematical bifurcation also
appears to be evident in the Putman \etal\ (2003; Fig.~11) dataset,
particularly over the range \mbox{$-30\deg < \theta < +10\deg$},
comparable both spatially and kinematically with that seen in
Figure~\ref{fig:vlsrtheta} here.  This bifurcation requires further
analysis and will be the subject of Paper~II of this series.

\section{Discussion and Conclusion}

The combined neutral hydrogen gas mass in the Magellanic Clouds,
Stream, Leading Arm, and inter-Cloud region, is in excess of $10^9
\msun$, within a factor of three or so of the \HI{} mass of the Milky
Way itself.  Within the framework of hierarchical clustering, this
represents a significant reservoir of potential fuel for future
generations of star formation.  Our aim is to properly model the
formation, evolution, and ultimate fate of the gas (and stars)
associated with the Magellanic System.

The past decade has seen a wealth of new observational data for the
System appear in the literature, in addition to the benefits of
Moore's Law currently governing the increase in computational power.
In combination, the two have allowed us to formulate improved models
for the Magellanic Stream, a ``Rosetta Stone'' for the hierarchical
clustering scenario of galaxy formation.  We have presented here a
simulation with $\sim 40$ times the resolution of previous simulations
of the Stream, enabling us to examine more subtle features
(kinematically and spatially) not previously considered in the models.
Our gross results parallel those of Gardiner \& Noguchi (1996) and
Yoshizawa \& Noguchi (2003), partly by construction, but the improved
resolution and parameter space coverage here are unique.  The gross
features of both the trailing Stream and Leading Arm Feature are
successfully recovered.

The bifurcation of the Stream observed both spatially (Morras 1983)
and kinematically (Putman \etal\ 2003) had been previously suggested
to be due to the ``twisting'' motion of the SMC's orbit about the LMC.
Our simulations are consistent with this picture, with the thinnest
part of the Stream corresponding to the location where the orbits
cross at \mbox{$(l,b) \sim (45\deg,80\deg)$}. The presence of this
helix-like structure is an important test for any simulation that
wishes to model the Magellanic System.

Despite the successes of the model, the comparisons we have made to
date have uncovered several unresolved problems. First, there is still
too little flux in the modelled MS, with the observed Stream (within
the MSI--VI clumps) having 3--4 times (Putman \etal\ 2003) more mass
than the modelled stream. Second, both the MS and LAF are too ``long''
in the models, with the LAF also being somewhat displaced from that
observed.  Third, the ratio of LAF-to-MS gas mass appears to be too
high in the simulations, by a factor of at least 3--4.  Fourth, while
the LAF seems to have the correct deviation angle (cf. Gardiner 1999)
relative to the great circle traced by the MS, its origin is somewhat
offset from that inferred by the Putman et~al. (1998) dataset.
Finally, the velocity dispersion of the currently modelled inter-Cloud
region remains too high, but we speculate that the inclusion of gas
dissipation in our future studies may alleviate this discrepancy.

Paper~II of this series will contain the full details of the parameter
space coverage undertaken in our work, as well as a thorough
examination of the spatial and kinematical bifurcations alluded to in
Section~3.  Paper~III incorporates gas dynamics, star formation,
radiative cooling, feedback, and chemical enrichment throughout the
Magellanic System.  We will re-examine the orbits of the Clouds
coupled with improved potentials for the LMC and Galaxy, as additional
data becomes available (particularly on the shape of the Galactic
potential -- e.g.  Martinez-Delgado \etal\ 2003; Bellazzini 2003; Helmi
2003).  A drag force term will also be introduced into the model, akin
to that adopted by Gardiner (1999), in an attempt to better reproduce
the geometry of the LAF.

The tools employed in analysing the simulations described here will
shortly be released to the public -- this software package affords any
user the ability to project virtually any N-body simulation into
various projections representative of the observer's ``plane'',
including the production of FITS files suitable for further analysis
by any other astronomical software package.

\section*{Acknowledgements}

We would like to thank Masafumi Noguchi who kindly provided his code
for the orbit calculation, and Mary Putman for the HIPASS HVC
cubes. TC would like to thank Chris Thom and Virginia Kilborn for
helpful discussions.  The support of the Australian Research Council
and the Victorian Partnership for Advanced Computing, the latter
through its Expertise Grant Program, is gratefully acknowledged.

\section*{References}

\reference Barnes, D.~G., Staveley-Smith, L., de~Blok, W.~J.~G.,
et~al., 2001, MNRAS, 322, 486
\reference Bellazzini, M., 2003, MNRAS, in press (astro-ph/0309312)
\reference Dieter, N.~H., 1965, AJ, 70, 552 
\reference Gardiner, L.~T. 1999, in Gibson, B.~K., Putman, M.~E. eds., 
Stromlo Workshop on High-Velocity Clouds, ASP, San~Francisco, p.~292 
\reference Gardiner, L.~T., Noguchi, M., 1996, MNRAS, 278, 191 
\reference Gardiner, L.~T., Sawa, T., Fujimoto, M., 1994, MNRAS, 266, 567 
\reference Gibson, B.~K., Giroux, M.~L., Penton, S.~V., Putman, M.~E., 
Stocke, J.~T., Shull, J.~M., 2000, AJ, 120, 1830
\reference Hardy, E., Suntzeff, N.~B., Azzopardi, M., 1989, ApJ, 344, 210 
\reference Harris, J., Zaritsky, D., 2001, ApJS, 136, 25
\reference Heller, P., Rohlfs, K., 1994, A\&A, 291, 743 
\reference Helmi, A., 2003, MNRAS, submitted (astro-ph/0309579)
\reference Ibata, R.~A., Gilmore, G., Irwin, M.~J., 1994, Nature, 370, 194
\reference Jones, B.~F., Klemola, A.~R., Lin, D.~N.~C., 1994, AJ, 107, 1333 
\reference Kawata, D., Gibson, B.~K., 2003a, MNRAS, 340, 908 
\reference Kawata, D., Gibson, B.~K., 2003b, MNRAS, 346, 135 
\reference Kuijken, K., Dubinski, J., 1995, MNRAS, 277, 1341 
\reference Kunkel, W.~E., Demers, S., Irwin, M.~J., Albert, L., 1997, ApJ,
488, L129
\reference Lin, D.~N.~C., Lynden-Bell, D., 1982, MNRAS, 198, 707 
\reference Lin, D.~N.~C., Jones, B.~F., Klemola, A.~R., 1995, ApJ, 439, 652 
\reference Lu, L., Sargent, W.~L.~W., Savage, B.~D., Wakker, B.~P.,
Sembach, K.~R., Oosterloo, T.~A., 1998, AJ, 115, 162  
\reference Maddison, S.~T., Kawata, D., Gibson, B.~K., 2002, Ap\&SS, 281, 421 
\reference Martin, N.~F., Ibata, R.~A., Bellazzini, M., Irwin, M.~J., 
Lewis, G.~F., Dehnen, W., 2003, MNRAS, in press (astro-ph/0311119)
\reference Martinez-Delgado, D., Gomez-Flechoso, M.~A., Aparicio, A.,
Carrera, R., 2003, ApJ, submitted (astro-ph/0308009) 
\reference Mastropietro, C., Moore, B., Mayer, L., Stadel, J.,
Wadsley, J. 2004, in Prada, F., Martinez-Delgado, D., Mahoney, T., eds.,
Satellite and Tidal Streams, ASP, San~Francisco, in press
(astro-ph/0309244)
\reference Mathewson, D.~S., Cleary, M.~N., Murray, J.~D., 1974, ApJ, 190, 291 
\reference Mathewson, D.~S., Schwarz, M.~P., Murray, J.~D., 1977, ApJ, 217, L5 
\reference Moore, B., Davis, M., 1994, MNRAS, 270, 209 
\reference Morras, R., 1983, AJ, 88, 62 
\reference Murai, T., Fujimoto, M., 1980, PASJ, 32, 581 
\reference Murali, C., Katz, N., Hernquist, L., Weinberg, D.~H., Dav\'e, R.,
2002, ApJ, 571, 1
\reference Putman, M.~E., 2000, PASA, 17, 1 
\reference Putman, M.~E., Gibson, B.~K., Staveley-Smith, L., et~al., 
1998, Nature, 394, 752 
\reference Putman, M.~E., Staveley-Smith, L., Freeman, K.~C., Gibson,
B.~K., Barnes, D.~G., 2003a, ApJ, 586, 170
\reference Putman, M.~E., Bland-Hawthorn, J., Veilleux, S., Gibson, B.~K.,
Freeman, K.~C., Maloney, P.~R., 2003b, ApJ, 597, 948
\reference Schommer, R.~A., Suntzeff, N.~B., Olszewski, E.~W., Harris,
H.~C., 1992, AJ, 103, 447 
\reference Smecker-Hane, T.~A., Cole, A.~A., Gallagher, J.~S., Stetson,
P.~B., 2002, ApJ, 566, 239 
\reference Spergel, D.~N., Verde, L., Peiris, H.~V., et~al., 
2003, ApJS, 148, 175
\reference Tolstoy, E., Venn, K.~A., Shetrone, M., et~al. 2003, AJ, 125, 707
\reference van der Marel, R.~P., Alves, D.~R., Hardy, E., Suntzeff, N.~B.,
2002, AJ, 124, 2639  
\reference van~Kuilenburg, J., 1972, A\&A, 16, 276
\reference Wannier, P., Wrixon, G.~T., 1972, ApJ, 173, L119 
\reference Weinberg, M.~D., 2000, ApJ, 532, 922 
\reference Weiner, B.~J., Williams, T.~B., 1996, AJ, 111, 1156 
\reference Yoshizawa, A.~M., Noguchi, M., 2003, MNRAS, 339, 1135 

\end{document}